\documentclass[twocolumn]{aastex631}
\usepackage{ amssymb }
\usepackage{amsmath}

\begin{document}

\title{Radiatively Active Clouds and Magnetic Effects Explored in a Grid of Hot Jupiter GCMs}
\correspondingauthor{Thomas Kennedy}
\email{thomak@umich.edu}

\author[0000-0002-2984-3250]{Thomas D. Kennedy}
\affiliation{Department of Astronomy and Astrophysics, University of Michigan, Ann Arbor, MI, 48109, USA}

\author[0000-0003-3963-9672]{Emily Rauscher}
\affiliation{Department of Astronomy and Astrophysics, University of Michigan, Ann Arbor, MI, 48109, USA}

\author[0000-0003-0217-3880]{Isaac Malsky}
\affiliation{Department of Astronomy and Astrophysics, University of Michigan, Ann Arbor, MI, 48109, USA}

\author[0000-0001-8206-2165]{Michael T. Roman}
\affiliation{School of Physics and Astronomy, University of Leicester, Leicester, LE1 7RH, UK}

\author[0000-0002-6980-052X]{Hayley Beltz}
\affiliation{Department of Astronomy, University of Maryland, College Park, MD 20742, USA}




\begin{abstract}

Cloud formation and magnetic effects are both expected to significantly impact the structures and observable properties of hot Jupiter atmospheres. For some hot Jupiters, thermal ionization and condensation can coexist in a single atmosphere, and both processes are important. We present a grid of general circulation models across a wide range of irradiation temperatures with and without incorporating the effects of magnetism and cloud formation to investigate how these processes work in tandem. We find that clouds are present in the atmosphere at all modeled irradiation temperatures, while magnetic effects are negligible for planets with irradiation temperatures cooler than 2000 K. At and above this threshold, clouds and magnetic fields shape atmospheres together, with mutual feedback. Models that include magnetism, through their influence on the temperature structure, produce more longitudinally symmetric dayside cloud coverage and more equatorially concentrated clouds on the nightside and morning terminator. To indicate how these processes would affect observables, we generate bolometric thermal and reflected phase curves from these models. The combination of clouds and magnetic effects increases thermal phase curve amplitudes and decreases peak offsets more than either process does individually.

\end{abstract}



\section{Introduction} \label{sec:intro}
Since the first discoveries of planets outside our solar system, hot Jupiters have emerged as a prominent class of planet that has no Solar System analog. With equilibrium temperatures in excess of 1000 K, they exist in a regime where irradiation from the host dominates the energy budget of the planet. Due to their large radii, high temperatures, and short orbital periods, they are the most detectable type of planet and the most amenable to atmospheric characterization. Observed hot Jupiters span a large range in irradiation temperature, the most important factor in setting the atmospheric conditions of the planet. Additionally, they span a range of compositions, rotation rates, and surface gravities, all of which contribute meaningfully to their structure. These factors combine to make hot Jupiters a diverse population of observationally accessible exoplanets.

Additionally, because the planets have orbital periods on the order of days, tidal synchronization is presumed to be nearly universal among hot Jupiters \citep{Rasio1996}. This places hot Jupiters, and particularly their atmospheres, into an inherently three-dimensional regime, with a permanent dayside and nightside. As a consequence of this, three-dimensional General Circulation Models (GCMs) have been and continue to be a crucial tool for understanding how physical processes shape these extreme systems.
GCMs numerically solve the fluid dynamics equations for the atmosphere, coupled to some approximation of radiative transfer, to find a quasi-equilibrium structure. Despite the variety of numerical treatments and approximations employed, GCMs show agreement on the broad strokes of hot Jupiter dynamics \citep[see][for a recent review]{Showman2020review}. Global-scale patterns develop, most notably an eastward equatorial jet with wind speeds of a few km s$^{-1}$ and strong day-night temperature contrasts. Depending on how efficiently the atmosphere re-radiates heat, this can shift the hottest point of the planet eastward of the substellar point. The eastward-shifted hotspot, first predicted by early GCMs \citep{Showman2002, Cooper2005,DobbsDixon2008}, has now been observed in many hot Jupiters across the population \citep{Beatty2019, Keating2019, Bell2021, May2022}.

GCMs have frequently been used to interpret observations of individual planets \citep[e.g.][]{Showman2009,Lewis2014hatp2b,Roman2017,Drummond2018HD209,Mendonca2018W43b, Beltz2021HD209hires,Teinturier2024W43b}, but have also been used to understand how physical processes drive trends across the population. This is usually done with a grid of GCMs spanning the hot Jupiter parameter space in irradiation, with the addition of variation in metallicity, rotation rate, and/or gravity \citep[][]{Perna2012grid,Kataria2016grid,Komacek2017,Komacek2020variabilitygrid,Parmentier2021grid,Baeyens2021grid,Roman2021,Komacek2022tintgrid,Tan2024GCMgrid,Roth2024grid}.

While all GCMs model fluid dynamics and radiative transfer with some level of simplifying assumptions, other physical processes are expected to be important in shaping the structure and observable characteristics of hot Jupiters and have been studied in GCMs. These include disequilibrium chemistry \citep[e.g.][]{Steinrueck2019diseqGCM,Zamyatina2023diseq, Lee2023}, hydrogen dissociation and recombination \citep[e.g.][]{Tan2019UHJcirc, Tan2024GCMgrid}, condensate clouds \citep[e.g.][]{Lee2016cloudGCM,Lines2018cloudGCM,Roman2019}, photochemical hazes \citep[e.g][]{Steinrueck2021hazeGCM}, and magnetohydrodynamic (MHD) effects \citep[e.g.][]{Rauscher2013, Rogers2017}. There is no GCM that can currently model all of these processes in tandem, even with simple prescriptions. It is probable, however, that these processes can combine in non-linear ways. To move toward a holistic understanding of hot Jupiter atmospheres, therefore, it is useful to model these processes jointly.

In this work, we focus on the condensation of clouds and MHD effects. When conditions are cool enough at a given pressure, gaseous species condense into clouds. Because of this temperature dependence, clouds are expected to be most important in cooler planets and cool regions (e.g. nightsides) of hotter planets. Once clouds are present, they act as an additional source of absorption and scattering, altering the atmosphere's thermal structure, muting spectral features, and reflecting some starlight back to space. MHD effects become important when parts of the atmosphere become thermally ionized and interact with the planetary magnetic field. As such, these effects will be most important for daysides of hotter planets. When non-negligible, The first-order effect of MHD is a spatially variable and directional drag on the winds, which in turn modifies the temperature structure of the atmosphere. These processes have been modeled individually, but never together.

Here, we present a grid of hot Jupiter GCMs with irradiation temperatures\footnote{{Irradiation temperature is defined as T$_{\rm irr} = T_* \sqrt{\frac{R_*}{a}}$, where T$_*$ is the stellar effective temperature, $R_*$ is the stellar radius, and $a$ is the semimajor axis of the planet's orbit. It differs from equilibrium temperature by $T_{\rm irr}=\sqrt{2}\,T_{\rm eq}$.}} ranging from 1000 K to 3250 K, with and without magnetic effects and condensate cloud formation to investigate how these processes act in tandem across the hot Jupiter population. Both prescriptions are ``active," meaning that they affect the atmospheric structure as the simulation progresses and can therefore have feedback with one another. We present a summary of our GCM (including cloud and magnetic drag prescriptions) and the chosen model parameters in Section \ref{sec:methods}. Results of our model grid are presented in Section \ref{sec:results}. A discussion of how modeling choices may affect our results is presented in Section \ref{sec:discussion}, and our conclusions are summarized in Section \ref{sec:conclusions}.

\section{Methods}
\label{sec:methods}
The atmospheric models presented here were run using the RM-GCM version 5.0, which is open-source and available on Github \citep{RM-GCM}. The RM-GCM was adapted from an Earth atmospheric modeling code \citep[the University of Reading's Intermediate General Circulation Model,][]{Hoskins1975} for the hot Jupiter regime by \citet{Rauscher2010}. Since then, functionality has been added for double-gray radiative transfer \citep{Rauscher2012}, a prescription for magnetic drag \citep{Rauscher2013}, radiatively active clouds \citep{Roman2017,Roman2019,Roman2021}, and picket fence radiative transfer \citep{Malsky2024}. The models presented here use the RM-GCM version presented in \citet{Malsky2024}, and summarized here. The dynamical core employed by the RM-GCM solves the primitive equations of meteorology in spectral space  horizontally, and vertically on a grid logarithmically spaced in pressure (specifically, $\sigma =$ P/P$_s$).

\subsection{Picket-fence Radiative Transfer}
All of the models presented here use the two-stream ``picket-fence'' radiative transfer method implemented in \citet{Malsky2024}. Originally developed for stellar atmospheres as a simple way to parameterize the effects of line blanketing \citep{Chandrasekhar1935, Mihalas1978}, picket-fence was extended to irradiated atmospheres by \citet{Parmentier2014, Parmentier2015}.
The method uses five opacity bands, two tracking thermal emission by the planet and three tracking starlight. Both absorption and scattering are included in our two-stream treatment in all bands, following the approach of \citet{Toon1989}. The two thermal channels represent continuum and line/band opacity in planet-emitted light, and the three starlight channels each capturing a third of the starlight absorbed by the atmosphere. Each channel has a set opacity in units of the Rosseland mean opacity, which is in turn taken from the pressure- and temperature-dependent fits of \citet{Freedman2014} for each part of the atmosphere.
This approach offers a simplified way to capture the multiwavelength effects of lines and bands. \citet{Lee2021} conducted a comparison between GCMs of HD\,209458b using correlated-k, picket-fence, and double-gray radiative transfer, finding that picket-fence results compare favorably with those of the more complex correlated-k approach with less computational cost (and were an improvement upon the double-gray scheme). In all of the models presented here, we utilize the coefficients derived by \citet{Parmentier2015} for equilibrium chemistry (i.e. without artificially removing TiO/VO from the atmosphere to mimic the effects of cold-trapping) for planets orbiting G-type stars.

\subsection{Radiatively Active Clouds}
The core of the cloud modeling used by the RM-GCM is detailed in \citet{Roman2019}, with the updates described in \citet{Roman2021} and \citet{Malsky2024}. Clouds are modeled as temperature-dependent sources of extinction and scattering that form and dissipate actively as the simulation progresses. This has the advantage of allowing clouds to actively shape the temperature and wind structure, which in turn shape the cloud distributions. This advantage is crucial for characterizing feedback processes within a GCM.

At each radiative timestep of the simulation, local temperature-pressure conditions are evaluated against a pre-computed condensation curve for each cloud species from \cite{Mbarek2016condcurves}. If the temperature is cooler than the condensation curve at the relevant pressure, a cloud forms.
A limit (in units of vertical layers) can be placed on the thickness of the clouds in each profile. A cloud is truncated a chosen number of layers above the lowest point in a given column of atmosphere where it condenses. This is evaluated for each cloud species independently. Physically, this represents a balance between particle size-dependent gravitational settling, vertical mixing, and other microphysical processes.
The strengths of vertical mixing \citep{Moses2013,Agundez2014} and sedimentation efficiency \citep{Christie2021mixing_sedimentation} are poorly constrained in hot Jupiter atmospheres, so the best choice of cloud extent in our framework is unclear. For our fiducial cloudy case, we choose to allow condensation anywhere local T-P conditions permit, without limiting vertical cloud thickness \citep[equivalent to the ``extended'' cases of ][]{Roman2021,Malsky2024}. This assumes strong vertical mixing, and should set an upper limit on the influence of clouds on hot Jupiter atmospheres. The choice of maximum cloud extent is most impactful in the coolest models (with deeper cloud-bases). For the cooler planets we model in this work (T$_\mathrm{irr}\leq 2000$\,K), therefore, we run additional model versions with more vertically compact cloud distributions. {In all models, cloud particle size is prescribed as a function of pressure as in \citet{Roman2021} \citep[based on the results of ][]{Lines2018cloudGCM}, fixed to 0.1 $\mu$m for $P\leq 10$ mbar and {growing linearly in pressure} to $\sim 100\ \mu$m at 100 bar, as:
\begin{equation}
\rm{a}(\rm{P}) =
\begin{cases}
      0.1 \,\mu \rm{m} &  P \leq 10^{-2} \,\rm{bar} \\
      \big(0.1 + \frac{\rm{P}}{1\rm{bar}}\big) \,\mu \rm{m} & P > 10^{-2} \,\rm{bar}
\end{cases}
\end{equation}}

The RM-GCM has functionality to model thirteen cloud species, but five of these (ZnS, Na$_2$S, MnS, Ni, and Fe) are expected to have limited presence in hot Jupiter atmospheres due to high nucleation barriers \citep{Gao2020}. Consequently, in the GCMs presented here we only include the other eight: KCl, Cr, SiO$_2$, Mg$_2$SiO$_4$, VO, Ca$_2$SiO$_4$, CaTiO$_3$, and Al$_2$O$_3$. The atmosphere is assumed to be well-mixed, with the mole fractions of each cloud species set by the availability of the stoichiometrically-limiting atomic constituent (assuming solar abundances). {While some cloud-forming species share components, none of them share a limiting atom, so this is ignored for the purpose of setting abundances.} When the conditions described above allow a cloud to form in the RM-GCM, a set fraction  of the available cloud-forming mass condenses. {The appropriate value for this fraction is not well-constrained, and should in principle be set by how the partial pressure of the condensate compares to its equilibrium vapor pressure, the efficiency of rainout, and the availability of condensation nuclei \citep{Roman2019}. In this work, we assume that 10\% of the available cloud-forming mass condenses \citep[following][]{Roman2019,Roman2021,Malsky2024}. This is largely a practical choice, as allowing all of the available cloud-forming material to condense produces heating rates that challenge the numerical stability of our model.} If the temperature of a cloudy region rises above the condensation curve, the cloud dissipates. This approach assumes that cloud formation and dissipation occurs faster than the time it takes to advect across a grid cell. \citet{Powell2018} perform simulations of microphysical processes coupled to vertical transport, and show that cloud evaporation timescales are typically of order 1 s, while condensation takes $\gtrsim10^3$ s. Given this, instantaneous evaporation is a safe assumption with the horizontal resolution and typical wind speeds of the models presented here. Formation timescales are long enough for gas to advect a few percent of the planetary radius while forming, however, so we may overestimate cloud coverage, particularly on the evening limbs of hot planets.

When a cloud is present, it contributes to the opacities in each of the five channels. In the thermal channels, cloud opacities are the Rosseland (low-opacity channel) and Planck (high-opacity channel) mean opacities for the relevant cloud species. The three starlight channels use the cloud absorption and scattering properties evaluated at 500, 650, and 800 nm, respectively, chosen to span the bulk of emission by a G-type host.
{These properties are interpolated from a pressure-temperature tables (Rosseland and Planck means) or pressure-wavelength tables (discrete wavelengths), which were pre-computed using Mie theory. Due to our assumption of spherical particles of a single particle size in each layer, the extinction coefficient Q$_{\rm ext}$ can oscillate in wavelength. These oscillations can occur at starlight wavelengths in deeper regions of the atmosphere (P $>$ 0.01 bar), where particle sizes are comparable to starlight wavelengths. Sampling several wavelengths across our channels mitigates this to some extent, but not entirely. We do not expect this to be a dominant source of error in our radiative feedback}. When multiple cloud species overlap, the scattering properties are taken to be the optical depth-weighted averages of the optical properties of individual species \citep[as outlined in][]{Roman2019, Malsky2021}.

\subsection{Kinematic MHD}
As stellar irradiation increases across the hot Jupiter population, portions of the dayside will become hot enough for an appreciable fraction of the gas to thermally ionize. As the ions are embedded in the mostly-neutral flow, any parcel of atmosphere containing ions will experience a bulk Lorentz drag as it is advected across magnetic field lines.
In the RM-GCM, this interaction is parameterized using the ``kinematic MHD" approach introduced in \citet{Rauscher2013}. This approach assumes that the planetary magnetic field is a dipole aligned with the planet's rotation axis and that the flow of ions in the atmosphere {only induces a small perturbation on that background field. These assumptions allows us to parameterize the strength of Lorentz drag on the zonal winds} using a magnetic drag timescale, $\tau_{mag}$, as derived in \citet{Perna2010a}:
\begin{equation}
    \tau_{mag}(B, \rho, T, \phi) = \frac{4\pi\rho\,\eta(\rho,T)}{B^2|\text{sin}\phi|}.
\end{equation}
A global surface field strength B is prescribed, and magnetic resistivity $\eta$ is calculated as in \citet{Menou2012}:
\begin{equation}
    \eta = 230\sqrt{T}/\text{x}_e \,\,\, \text{cm}^2 \text{s}^{-1}
\end{equation}
With ionization fraction x$_e$ calculated using the Saha equation, including all elements from hydrogen to nickel. This drag is then applied to the zonal wind as an additional term in its derivative ($\frac{du}{dt} = -\frac{u}{\tau_{mag}}$). {For simplicity, drag on the meridional winds is ignored.} Any energy removed by this drag is returned to the atmosphere as local Ohmic heating. For numerical stability, we set a minimum drag timescale of one two-hundredth of a planet day (approximately 30 dynamical timesteps). This drag is evaluated locally, allowing for orders of magnitude variation in strength from the hot dayside to the cooler nightside, and at every dynamical timestep, allowing for feedback with the temperature structure of the planet.

{The assumption that the field induced by the ions in the atmosphere is only a small perturbation on the background magnetic field is a safe one if the magnetic Reynolds number R$_m < 1$. If R$_m$ exceeds unity, the atmosphere can couple to the magnetic field strongly enough to sustain a field via dynamo action, in which case magnetic effects on the flow would be complicated significantly \citep[e.g.][]{Rogers2014}. For all of our models where magnetic drag impacts the circulation (e.g., for T$_{\rm irr} > 2000$ K), we find regions of the atmosphere where R$_{\rm m}$ exceeds 1, albeit to different extents. 
For the T$_{irr} = 2000$ K models, there are only a few regions of the dayside atmosphere where R$_{\rm m}$ is marginally greater than 1, but in the hottest models, practically the entire dayside has R$_{\rm m} > 1$.
This means that we are likely over-simplifying the effects of magnetism by not solving the induction equation fully \citep[e.g.][]{Rogers2017, Hindle2021MHD}. In order to achieve our science goal of studying magnetic effects on circulation with cloud radiative feedback across T$_{irr}$, we are computationally limited to a simplified prescription. In lieu of a computationally feasible, more physically realistic alternative, this temperature-dependent, physically motivated approach is a plausible way to diagnose how population-level trends might be impacted by magnetic effects to first order.}

While magnetic fields of exoplanets have not yet been measured directly, dynamo modeling suggests that hot Jupiters should have field strengths of anywhere from 10-250 G depending on mass, age, and irradiation \citep{Reiners2010, Yadav2017}. Previous work using this kinematic MHD framework in the RM-GCM found a best match to observations of WASP-76b using a 3 G field strength after testing 0, 0.3, 3, and 30 G \citep{Beltz_structure}. Field strengths in excess of 30 G introduce numerical instabilities to our model in the hottest planets. Additionally, the transition from a non-dragged to magnetically dragged circulation is fairly stark; in our prescription, the dynamics of planets with 3 G and 10 G are far more similar than non-dragged and 3 G, so the precise field strength (above a few Gauss) will not be very impactful for a exploratory parameter study like this one. For these reasons, we chose a field strength of 3 G for all of our magnetically dragged models.

\subsection{Model Grid}
We ran GCMs for irradiation temperatures from 1000 K to 3250 K in steps of 250 K (10 irradiation temperatures). In order to isolate the causes of the trends identified, 4 base models were run for each T$_\mathrm{irr}$; one with neither clouds nor magnetic drag, one with magnetic drag but no clouds, one with clouds but no magnetic drag, and one with both clouds and magnetic drag, for a total of 40 base models. When magnetic drag is included, we assume a dipolar field with a surface strength of 3 G. When clouds are included, we allow 8 species to condense (KCl, Cr, SiO$_2$, Mg$_2$SiO$_4$, VO, Ca$_2$SiO$_4$, CaTiO$_3$, and Al$_2$O$_3$) with no imposed limit on vertical thickness. For the least-irradiated models (T$_\mathrm{irr}\leq$ 2000 K), we ran two additional cloudy versions with a maximum thickness of $\sim3$ and $\sim$9 pressure scale heights (10 additional models). Besides magnetic field strength and cloud prescription, all other system and model domain parameters were chosen to represent an ``average" hot Jupiter, and are fixed across the grid (Table \ref{tab:grid_parameters}). Parameters that ensure numerical stability (hyperdissipation timescale, sponge layer timescale, dynamical \& radiative timesteps), are tuned for each value of T$_\mathrm{irr}$, and can be found in Table \ref{tab:appendix_table}. 

Models were initialized with no winds and a single global-average double-gray pressure-temperature profile \citep{Guillot2010}, and run for 2000 orbits, enough for all but the deepest layers to come into a quasi-equilibrium state. All models were run with T31 horizontal resolution (corresponding to 96 longitudes and 48 latitudes), and 50 vertical layers. Dynamical timesteps of $\sim$25 s were used for models with T$_\mathrm{irr} \leq 2000\,\text{K}$, $\sim$20 s for models with T$_\mathrm{irr}$ of 2250 and 2500 K, and $\sim$15 s for models with T$_\mathrm{irr} \geq 2750\,\text{K}$. 

\begin{deluxetable}{lll}
{\tablehead{\multicolumn{3}{c}{Fixed Grid Parameters}}}
\startdata
\multicolumn{1}{c}{Parameter} & \multicolumn{1}{c}{Value} & \multicolumn{1}{c}{Unit} \\
\hline
\hline
Gravitational Acceleration                             & 20                    & m s$^{-2}$        \\
Planet Radius                                          & 9.4$\times$10$^7$      & m                 \\
Internal Temperature (T$_\mathrm{int})$                         & 500                  & K                 \\
Rotation/Orbit Period & 1.4 & days \\
Planetary Rotation Rate                                & 5.19$\times$10$^{-5}$ & radians s$^{-1}$  \\
Metallicity & 1 & solar \\
Specific Gas Constant $\mathcal{R}$ & 3523 & J kg$^{-1}$ K$^{-1}$\\
R/c$_\mathrm{p}$ & 0.286 & \\
Base Pressure
       & 100 & bar \\
Top Pressure
       & 10$^{-4}$ & bar \\
Vertical layers
       & 50 &  \\
\hline
\enddata
\label{tab:grid_parameters}
\caption{The fixed parameters used across our grid of GCMs.}
\end{deluxetable}

\subsection{Bolometric phase curve generation}
We generate bolometric phase curves from each model, in both thermal emission and reflection. To generate the model phase curves {from a given model snapshot}, the total outgoing flux in the two thermal channels (thermal phase curves) or three starlight channels (reflected phase curves) of the GCM, $F_{out}$, is integrated across the visible hemisphere at a given phase, with the assumption that emission (or reflection) is isotropic at the top of the atmosphere. {Total outgoing flux toward a given sub-observer longitude $\Theta$ is then:}
\begin{equation}
    F(\Theta) = \frac{\sum_{\theta,\phi}F_{out}(\theta,\phi)\text{cos}^2(\phi)\text{cos}(\theta-\Theta)}{\sum_{\theta,\phi}\text{cos}^2(\phi)\text{cos}(\theta-\Theta)}
\end{equation}
{where $\theta$ and $\phi$ are the longitude and latitude of a given column, respectively. The sum is taken from $\Theta -90$ to $\Theta + 90$ degrees (e.g. the far side of the planet is omitted from the summation). Here we omit the R$^2 \Delta\phi\Delta\theta$ in both numerator and denominator, as the columns are equally spaced in latitude and longitude and we assume a constant radius.}
From the phase curves we can calculate two standard properties: the normalized amplitude ($[F_{\mathrm{max}}-F_{\mathrm{min}}]/F_{\mathrm{max}}$) and the offset of the planet's peak emission relative to secondary eclipse (with positive values indicating brighter regions to the east of the substellar point). We present both values for the thermal phase curves and offsets for the reflected light phase curves, together with Bond albedo values{, calculated as} 
\begin{equation}
    A_B = \frac{\sum_{\theta,\phi}{F_{\rm refl}}(\theta,\phi) \mathrm{cos}(\phi)}{\sum_{\theta,\phi}F_{\rm inc}(\theta,\phi)\mathrm{cos}(\phi)}
\end{equation}
{where F$_{\rm refl}$ is the upward top-of-atmosphere flux in the starlight channels and F$_{\rm inc}$ is the incident flux received from the host star (accounting for projection effects).}
While more detailed comparisons with observations would require wavelength-dependent radiative post-processing, these bolometric values give us a rough idea of how the global patterns predicted by the models would manifest observationally.

\section{Results}
\label{sec:results}
\begin{figure*}[h]
    \centering
    \includegraphics[width=\textwidth]{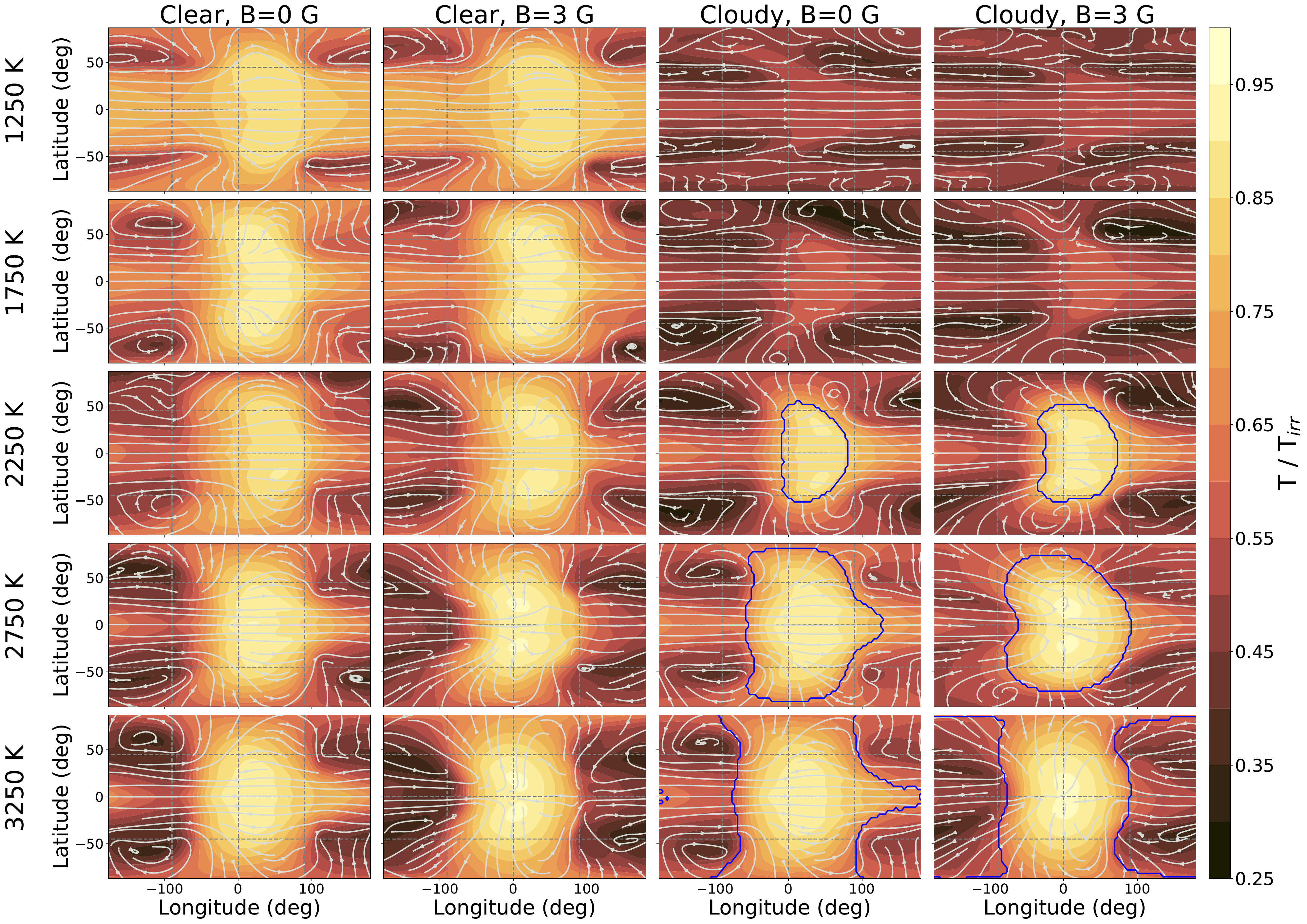}
    \caption{Normalized temperature (T/T$_\mathrm{irr}$) and wind maps at the 0.1 bar pressure level for a subset of the irradiation temperatures in our model grid, for each of the four base cases (clear and extended clouds, nonmagnetic and B = 3 G). The blue contour separates regions where IR photospheres are cloud-free from those where clouds alter the depth of the IR photosphere. Below T$_\mathrm{irr}$ = 2000 K, cloud coverage is global at $\sim0.1$ bar, and models are significantly cooled by reflection of starlight. Above T$_\mathrm{irr} = 2000$\,K, the bulk eastward flow is significantly dragged by magnetic effects and the hotspot is moved toward the substellar point relative to non-dragged models.}
    \label{fig:summary}
\end{figure*}

In the clear, non-magnetic models, we recover the broad trends in irradiation temperature shown in previous studies \citep{Perna2012grid,Roman2021,Komacek2017}. Figure \ref{fig:summary} shows temperature maps across the model grid at 0.1 bar (near the base of the IR photosphere) with wind patterns and a contour indicating cloud coverage overlaid for every other T$_\mathrm{irr}$ value modeled, with the leftmost column showing the clear, nonmagnetic models. A similar figure for the models omitted from this plot can be found in the Appendix (Figure \ref{fig:appendix_summary}). The dynamics are dominated by an eastward equatorial jet on the order of several km s$^{-1}$, with jet speed increasing as a function of irradiation temperature. As irradiation increases, so does the day-night contrast. This trend can be understood by comparing the relevant timescales \citep[e.g.][]{Showman2002, Perez-Becker2013}.
Day to night heat redistribution is fundamentally set by a contest between advection of heat and radiative loss of energy to space. While timescales relevant to advection of heat to the nightside (such as the time for the jet to traverse a hemisphere or the time for a gravity wave to propagate across a hemisphere) decrease as temperatures increase, the radiative timescale decreases far more sharply, so it is the variation in this timescale that sets the behavior across the population to first order \citep{Roth2024grid}.
While different components of the global circulation can control overall day-night contrast and longitudinal structure \citep{Hammond2021, Lewis2022}, for a set of (clear, non-magnetic) models with fixed rotation period like ours, both day-night heat redistribution and hotspot shift scale inversely and monotonically with irradiation \citep{Perna2012grid,Roman2021}.
This manifests observably as a decreasing peak offset and an increasing amplitude in thermal phase curves as a function of increasing irradiation temperature.

In comparing our clear, non-magnetic models to those with clouds and/or magnetism, we find that clouds are important across the entire model grid, while magnetic drag only matters for T$_\mathrm{irr} \gtrsim$ 2000 K. Motivated by this, we organize our results into the following two sections: first, cool planets with roughly global cloud coverage and negligible magnetic drag, and then hot planets with spatially inhomogeneous cloud coverage and dynamically significant magnetic drag. 

\subsection{Cool models (T$_\mathrm{irr}\leq$ 1750 K)}

In the fiducial cloud-free case, we recover the temperature and wind structure we have come to expect for hot Jupiter atmospheres, with the dynamics dominated by a strong (several km/s) eastward equatorial jet displacing the hottest region of the planet eastward of the substellar point. In this range of parameter-space, ionization fractions are too low (and thus the magnetic resistivity is too large) for magnetic drag to have an appreciable impact on atmospheric dynamics.

\begin{figure}
    \centering
    \includegraphics[width=\columnwidth]{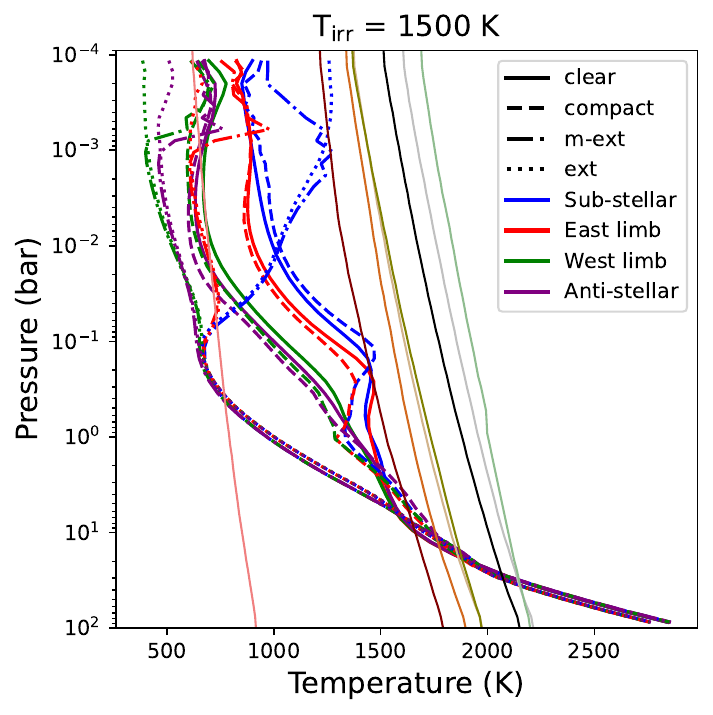}
    \caption{Meridionally-averaged temperature-pressure profiles for the substellar and antistellar longitudes and both terminators for models with T$_\mathrm{irr}=1500$\,K. Condensation curves for each of the 8 cloud species modeled are overlaid. The moderately-extended (``m-ext") cloudy case diverges from the fully-extended (``ext") model at the cloud top ($\sim10^{-3}$ bar), converging quickly to the corresponding clear structure. Compact cloudy models (``compact") only differ notably from clear models in the deep atmosphere ($\gtrsim10^{-1}$ bar),  where clouds are present. The condensation curves are, from coolest to hottest, KCl, Cr, SiO$_2$, Mg$_2$SiO$_4$, VO, Ca$_2$SiO$_4$, CaTiO$_3$, and Al$_2$O$_3$.}
    \label{fig:ext-clouds}
\end{figure}

The radiative impact of clouds is maximized in this regime, since most of the atmosphere is cool enough for a broad variety of cloud species to condense. In this regime, cloud condensation curves are crossed in the deep atmosphere, where the atmosphere at a given pressure is close to isothermal and the P-T profiles are shallow. This allows many condensation curves to be crossed in a narrow pressure range (see Figure \ref{fig:ext-clouds}). This results in the distributions of most cloud species being very similar to one another both horizontally and vertically. KCl, with a significantly cooler condensation curve, condenses higher than the other species, but has relatively little radiative impact due to the low abundance of K. The absorption and scattering of starlight higher than it would otherwise be deposited reduces equator to pole contrasts in the deep atmosphere ($\geq 1$ bar).  Simultaneously, increased thermal opacity from clouds makes cooling less efficient, resulting in greater heat redistribution in these layers. Together, these effects homogenize the deep atmosphere, leading to cloud distributions that are essentially one-dimensional, globally present above a deep cloud-base, regardless of species.
In contrast to the deep atmosphere's homogenization, the upper atmospheres of the extended and moderately-extended cloudy models have significantly greater day-night and equator to pole temperature contrasts than corresponding clear models (Figure \ref{fig:ext-clouds}). This is a consequence of starlight being absorbed higher in the atmosphere, where radiative timescales are shorter and heat redistribution is not as effective.

A consequence of the atmosphere being homogeneously cool enough for cloud formation is that the prescribed cloud thickness has a large impact on the distributions and radiative effects of clouds. A global cloud base, with a specified maximum cloud thickness, naturally leads to a global cloud-top at a fairly uniform pressure. Changing the cloud thickness, then, moves the cloud-top up and down in the atmosphere, but does not change the cloud bulk optical properties (set by the ratios of different constituents). This can cause discontinuities in the P-T profiles of ``moderately-extended" cloudy models, where the clouds are truncated high in the atmosphere. As all cloud species turn off in a few model layers, there is a precipitous drop in opacity in the upper atmosphere. The weak radiative coupling (low gaseous opacities) in these layers leads to the atmosphere above the cloud-top quickly snapping to a structure similar to that of a corresponding clear atmosphere (Figure \ref{fig:ext-clouds}). This shift happens high enough in the atmosphere ($\lesssim 1$ mbar) that emission and reflection from the planet is unaffected.

\begin{figure}
    \centering
    \includegraphics[width=\columnwidth]{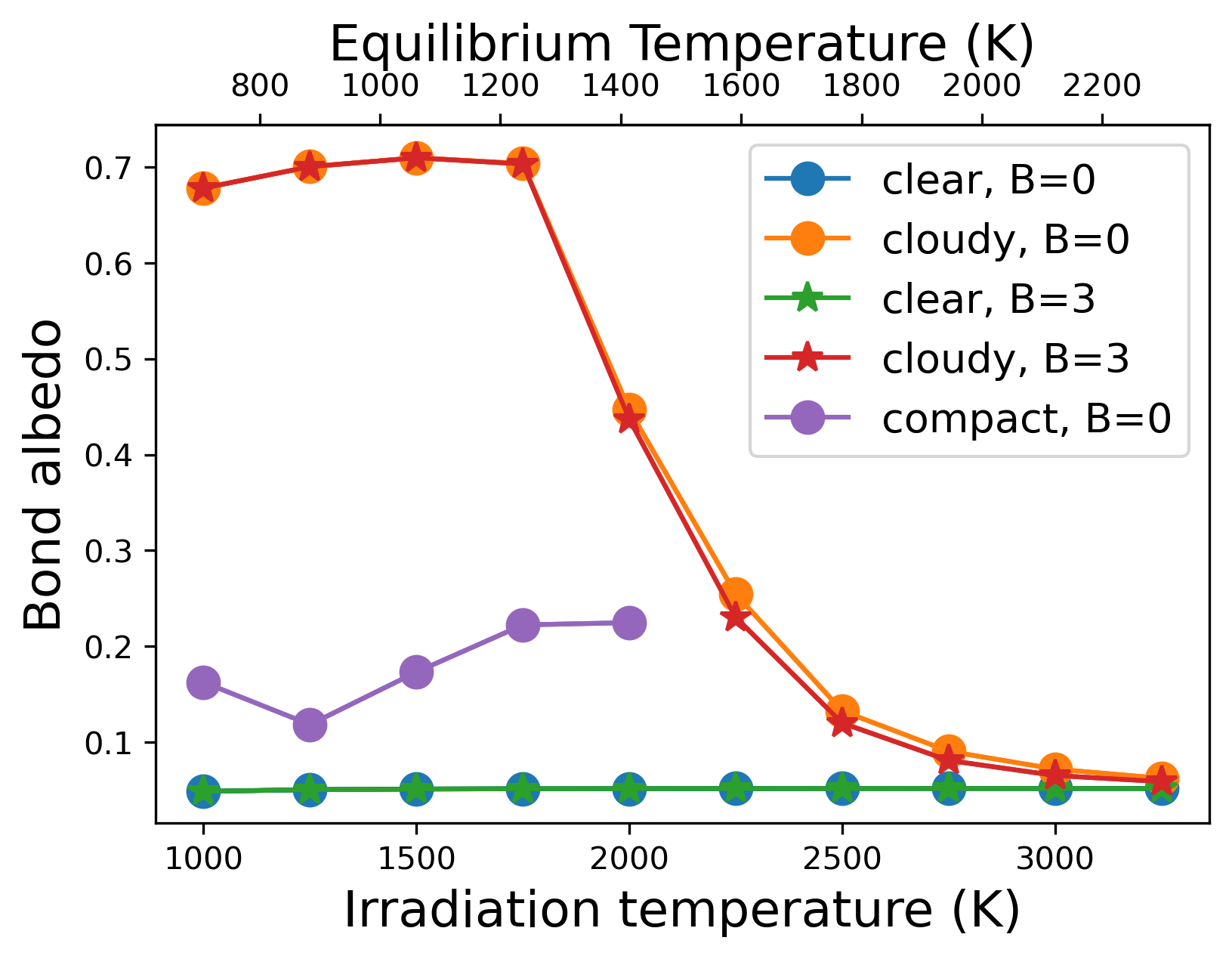}
    \caption{Bond albedos for each model case as a function of irradiation temperature. In clear models, a base albedo of 0.05 is prescribed to approximate the effect of Rayleigh scattering \citep[motivated by][see their appendix]{Malsky2024}. All other reflection is due to cloud scattering. In the extended (compact) cloudy models, albedos hover around 0.7 (0.2) until dayside clouds begin to dissipate at T$_\mathrm{irr} \geq 2000$\,K, after which point the two roughly align. Magnetic drag subtly decreases Bond albedo at intermediate and high irradiation temperatures by producing a more uniformly cloud-free dayside (see Section \ref{sec:cloudmag}).}
    \label{fig:albedo}
\end{figure}

{When all clouds are present, the optical properties are set mainly by SiO$_2$ and Mg$_2$SiO$_4$, as these two have the highest optical depth per bar. These silicate clouds are highly reflective, and as such} can increase the albedo of the planet significantly when present on the dayside. The magnitude of this effect is strongly dependent on the location of the cloud-top. Between the moderately- and fully-extended cloudy models, the difference is negligible, since little starlight is absorbed by the gas over the pressures between where the cloud-tops form for each case. For these models, the Bond albedo is frozen in at A$_B \sim 0.7$ until dayside clouds start to dissipate at T$_\mathrm{irr} \sim 2000$\,K \citep[consistent with the analogous ``extended nucleation-limited" models of][]{Roman2021}. In models with compact clouds, this is no longer the case, and the Bond albedo never exceeds about A$_B \sim 0.2$ (see Figure \ref{fig:albedo}). Compact cloudy models see a weak increase in albedo as a function of irradiation temperature. This is a result of the deep atmospheres becoming hotter, pushing the cloud base (and correspondingly the cloud top) to lower pressures.

\begin{figure*}
    \centering
    \includegraphics[width=\textwidth]{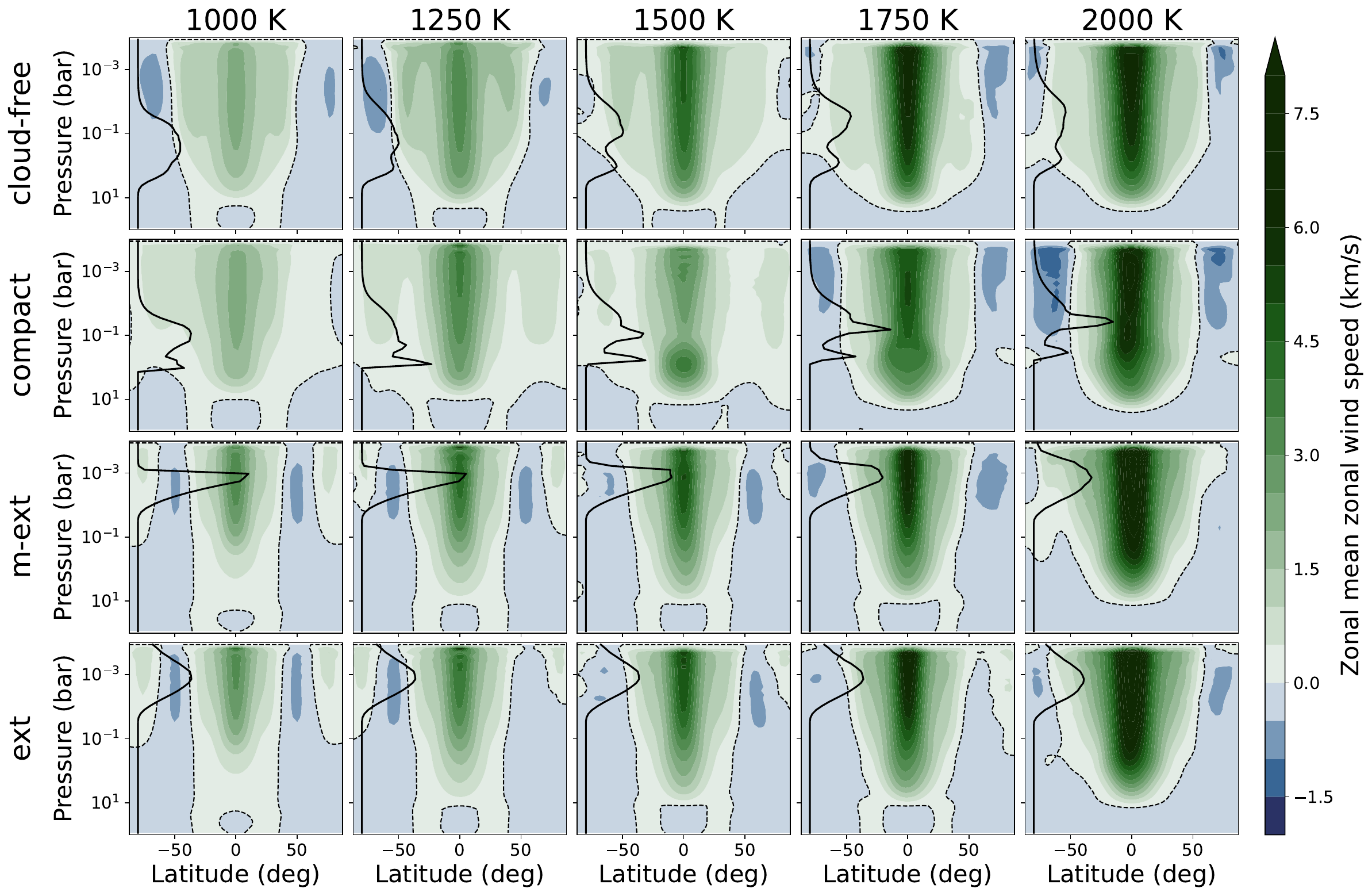}
    \caption{Zonally-averaged zonal mean wind speeds as a function of latitude and pressure (contours) for the cool half of our model grid. Eastward wind is positive. To give a sense of where starlight is deposited in each model case, the total starlight absorbed by each layer of the atmosphere is overlaid (black curve, scaled for visibility) . The strength, depth, and breadth of the equatorial jet are modified by the radiative feedback of clouds.}
    \label{fig:coolwinds}
\end{figure*}

The presence of clouds above the (clear) starlight photosphere strongly modifies where light is deposited, which in turn impacts the global dynamics. Figure \ref{fig:coolwinds} shows the zonal-mean zonal (eastward) winds as a function of latitude and pressure for models in this range, with the absorption of starlight (by both gas and clouds) overlaid as a function of pressure. As in previous work \citep[e.g.][]{Parmentier2021grid, Roman2021}, we find that the jet strength for any given model case increases as a function of irradiation temperature. For the extended and moderately-extended cloudy cases, the enhancement of upper atmosphere day-night and equator-pole temperature contrasts increases the maximum jet speed and shifts the base of the jet to higher altitudes. 

The radiative effects of compact clouds are more complex. In general, compact clouds broaden the jet at low pressures (above the cloud-top), but for some irradiation temperatures, compact clouds cause two eastward jets to form: one above the cloud deck and one below. 
Interpretations of compact cloudy models are further complicated by {localized heating of cloudy regions by the absorption of starlight, which can flatten or even invert dayside temperature profiles.}
Since the condensation curves of most species modeled here only span a few hundred Kelvin, when a cloud species induces a temperature inversion just above its base, some species (those with slightly hotter condensation temperatures) can be evaporated for almost the full range of their allowed thickness, limiting their influence on the atmosphere. In contrast to this, a local inversion can stop species with cooler condensation temperatures from condensing out in the deep atmosphere, and instead forming a second cloud deck at a higher altitude. To fully understand the range of possible multi-species cloud interactions with vertical limits would likely take a dedicated parameter exploration of its own.

\subsection{Hot models (T$_\mathrm{irr} \geq$ 2000 K)}
For T$_\mathrm{irr} \geq 2000$\,K, portions of the dayside atmosphere become ionized enough for magnetic drag to be significant with our assumed field strength of 3 G. The effects of magnetic drag on the circulation pattern are most pronounced in the upper atmosphere, where the dayside flow becomes almost entirely meridional, in agreement with previous work \citep[e.g.][Figure 3]{Beltz_structure}. Consequently, magnetically dragged models have poorer heat redistribution and hotspots progressively closer to the substellar point than corresponding non-dragged models. As in \citet{Beltz_structure}, we find that while our Lorentz drag prescription disrupts the dayside equatorial jet, the nightside still has a strong prevailing eastward wind (see Figure \ref{fig:summary}).

The onset of non-negligible thermal ionization coincides with the daysides being hot enough to begin dissipating clouds. Cloud coverage is inhomogeneous in this regime; as irradiation temperature increases, clouds first disappear from just east of the substellar point in the upper atmosphere, then progressively more and more of the dayside and deeper into the atmosphere. Clouds never entirely disappear from any column of the nightside in any model, but nightside clouds have limited presence in the deep atmosphere as nightside temperatures increase.

\subsubsection{Feedback between clouds and magnetic drag}
\label{sec:cloudmag}
\begin{figure*}[h]
    \centering
    \includegraphics[width=\textwidth]{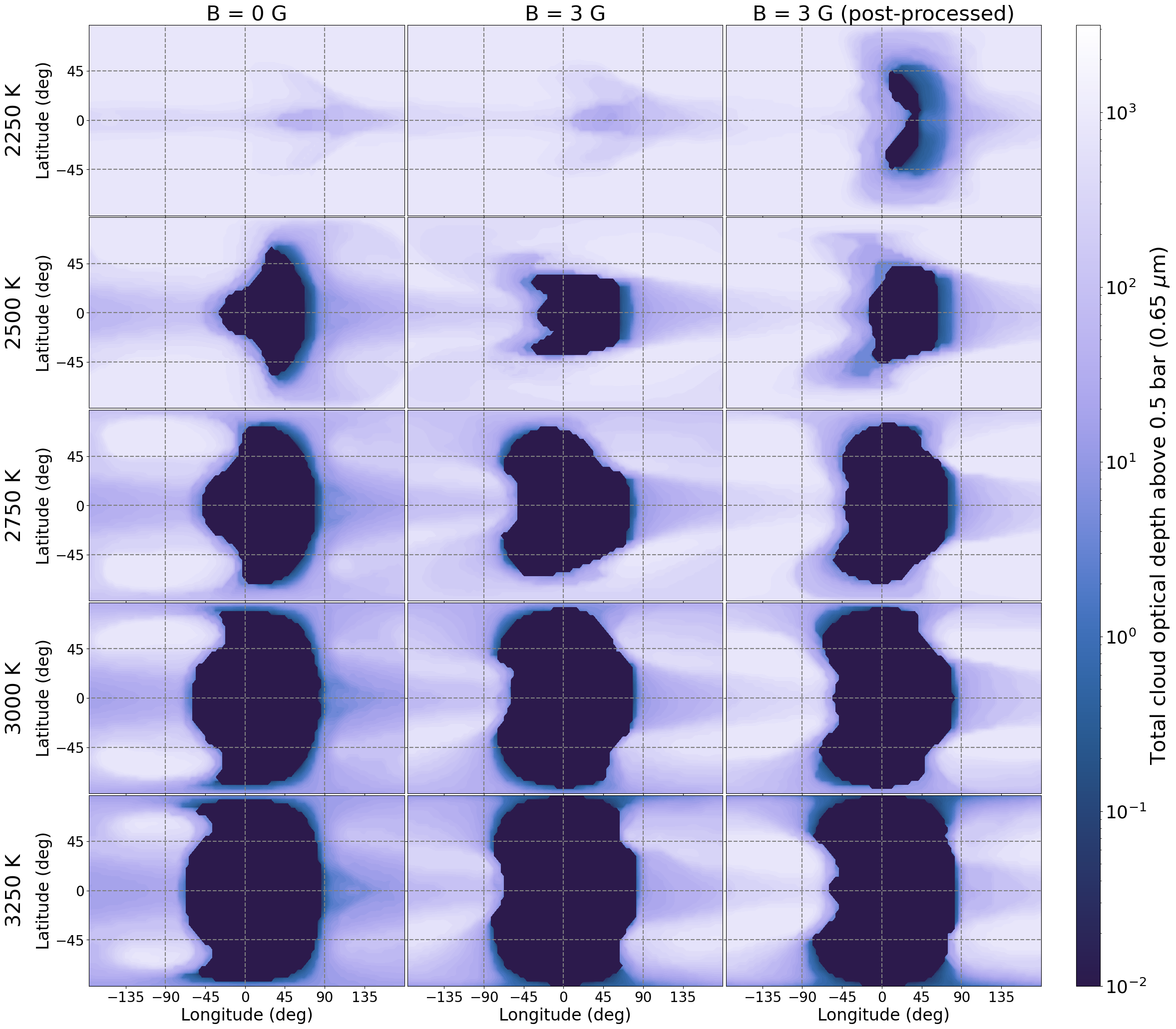}
    \caption{Maps of integrated cloud optical depth above 0.5 bar at 0.65 $\mu$m for various model cases at intermediate and high irradiation temperatures. Columns are: fully-extended cloudy models with no magnetic drag (left), fully-extended cloudy models with a 3 G magnetic field (middle), and a clear model with a 3 G magnetic field and post-processed clouds (right). Magnetically dragged models have more clouds to the east and fewer to the west on the dayside, as a result of the more symmetric temperature structure. Additionally, magnetic effects shift nightside and morning terminator cloud distributions toward the equator. Cloud radiative feedback itself also modifies the domain where clouds can persist, both by cooling the planet overall and by heating the deep nightside via the greenhouse effect (compare middle and right columns).}
    \label{fig:cloudcoverage}
\end{figure*}

This is the first time that actively forming and dissipating clouds with radiative feedback have been modeled in a GCM jointly with even a simplified temperature-dependent magnetic drag prescription. This provides us the first opportunity to explore the interaction between these two physical mechanisms and their joint effects on atmospheric structure. Figure \ref{fig:cloudcoverage} displays maps of total cloud optical depth (above 0.5 bar) across this range of irradiation temperatures for different model cases. Temperature structure (and therefore cloud distributions) are modified both by the inclusion of magnetic drag and by cloud radiative feedback. In light of this, we have displayed both fiducial cloudy cases (with and without magnetic effects) and a magnetically dragged model run without cloud radiative feedback, but with clouds post-processed on to the resulting model according to the same criteria applied in the GCM. A few key results appear.

First, once magnetic drag becomes effective across much of the dayside, the equatorial jet is damped and heat is instead primarily transported from the equator poleward. This produces more longitudinally symmetric dayside temperatures, which leads to correspondingly more symmetric cloud distributions (with cloudier evening terminators and fewer clouds west of the substellar point).

In addition to changing conditions on the planet's dayside, the magnetic disruption of the dayside equatorial jet results in a cooler nightside equator. Simultaneously, the more efficient polar heat transport produces significantly hotter nightside high-latitudes. Together, these effects shift nightside cloud distributions toward the equator. In non-dragged models, clouds are most significant in the nightside mid-latitudes. In dragged models, clouds are concentrated much closer to the equator. This cool air is then advected onto the morning limb by the eastward jet (still present on the nightside), producing more equatorial dayside cloud distributions compared to the higher-latitude distributions produced by non-dragged models.

Simultaneously as the magnetic effects are modifying cloud distributions, radiative feedback from the clouds themselves is also changing the picture (compare the middle and right columns of Figure \ref{fig:cloudcoverage}). For T$_\mathrm{irr}$ = 2250 K, the reflection provided by clouds cools the dayside enough to ensure more complete cloud coverage than would be produced by a model without cloud feedback. At higher T$_\mathrm{irr}$, clouds on the nightside inhibit the deep atmosphere from cooling, increasing temperatures and reducing cloud coverage at depth in the models with radiative feedback.

\begin{figure}
    \centering
    \includegraphics[width=\columnwidth]{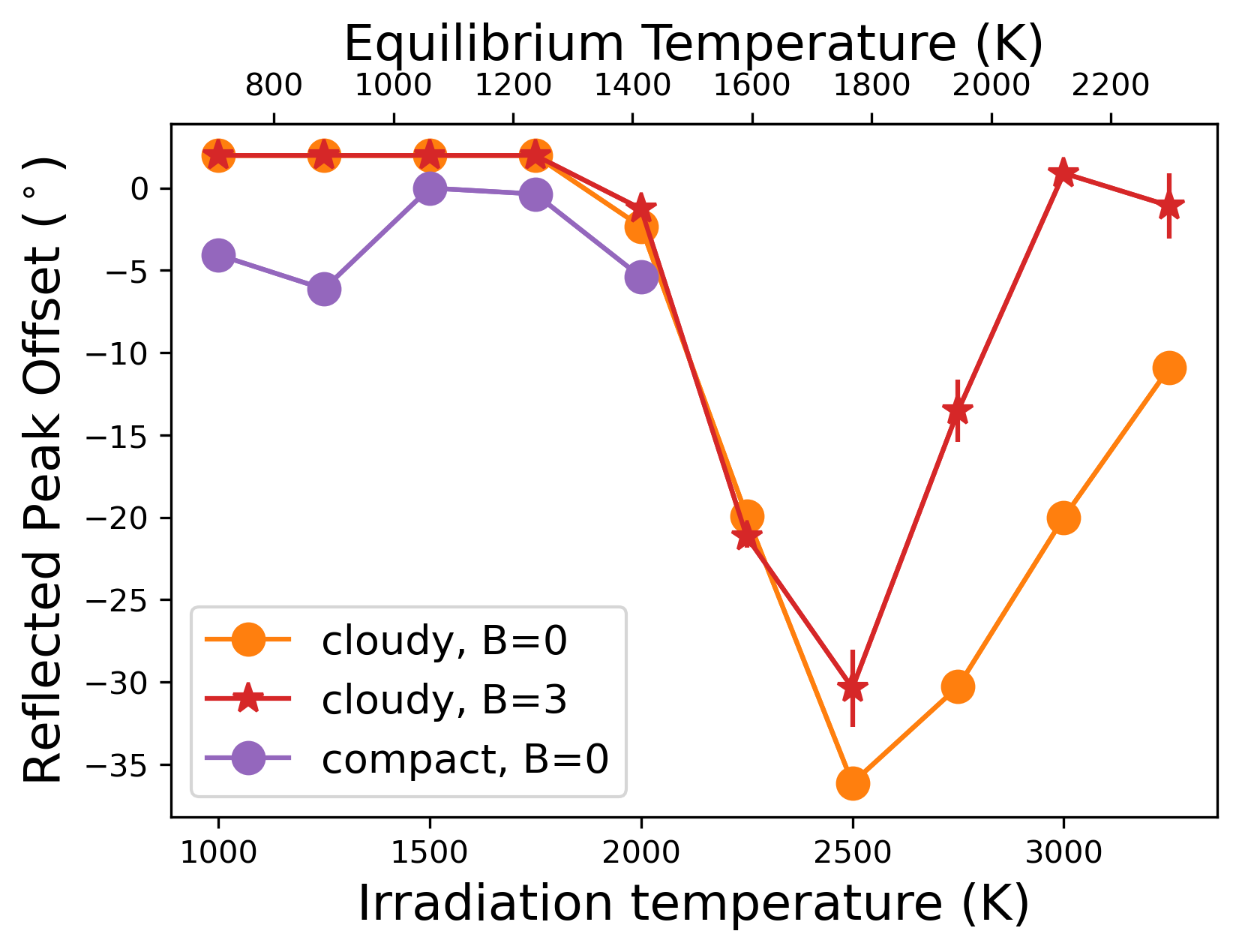}
    \caption{Offsets of reflected light phase curves for all cloudy models. For T$_\mathrm{irr} \geq$ 2500 K, the inclusion of magnetic drag limits cloud condensation west of the substellar point, reducing the westward shift in reflected light. Error bars indicate typical variability on scales of hundreds of planet days.}
    \label{fig:refle_pc}
\end{figure}

\subsection{Emitted and reflected light trends across T$_{\mathrm{irr}}$}

\begin{figure}
    \centering
    \includegraphics[width=\columnwidth]{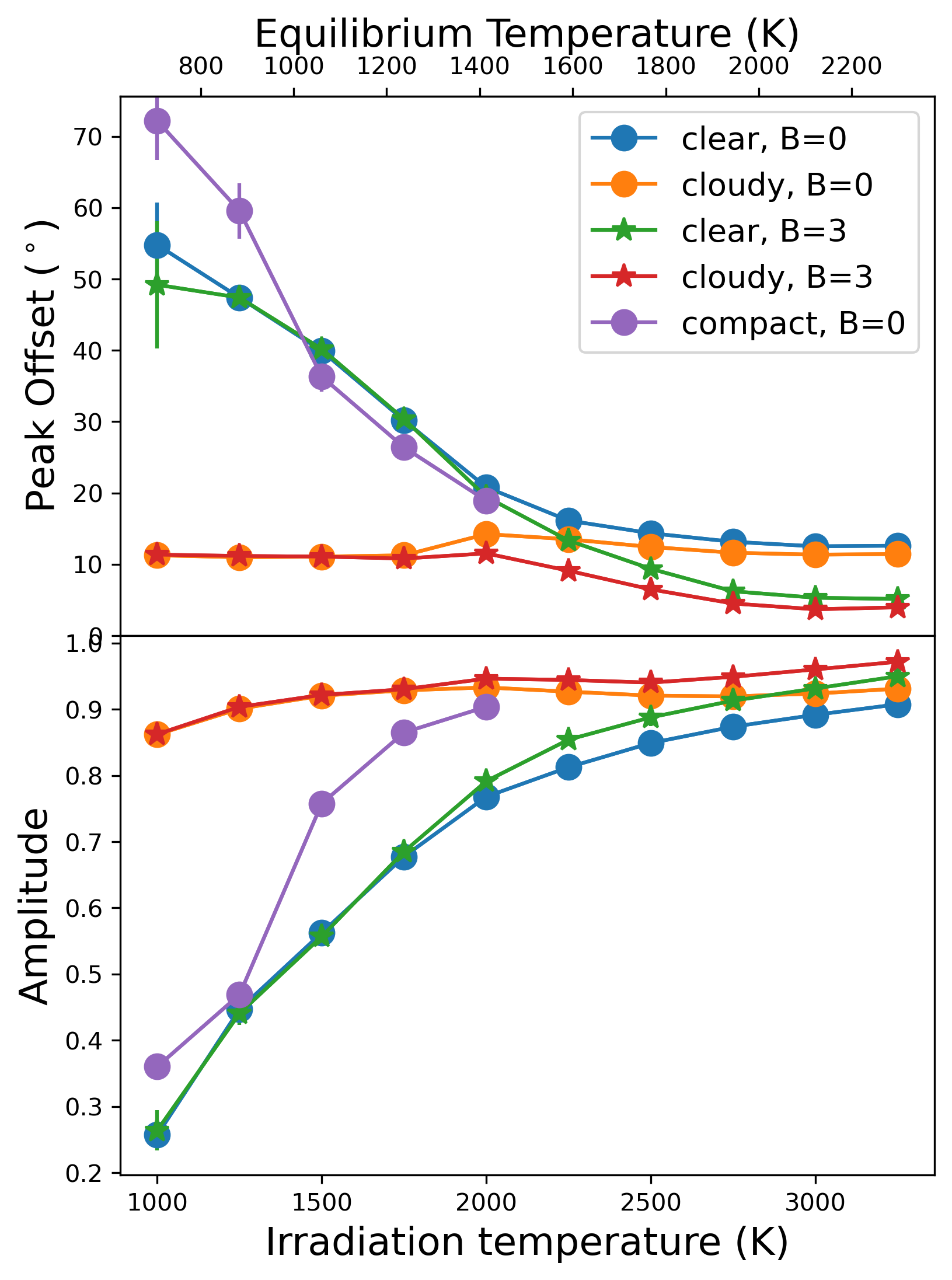}
    \caption{Bolometric thermal phase curve parameters for each model. The inclusion of either clouds, magnetic drag, or both strictly increases the amplitudes and decreases the offsets of phase curves. Peak offsets and amplitudes are predominantly set by clouds below T$_\mathrm{irr} = 2000$\,K, and both clouds and magnetic effects at and above that threshold. Error bars denote typical variability on scales of hundreds of planet days.}
    \label{fig:therm_pc}
\end{figure}

Figures \ref{fig:albedo}, \ref{fig:refle_pc}, and \ref{fig:therm_pc} provide a way of linking the physical trends discussed in the sections above to trends we might observe across the hot Jupiter population through the main parameters of emitted and reflected light phase curves.  As discussed at the beginning of the section, in clear, nonmagnetic models, the emitted phase curve amplitude increases and the offset decreases with increasing irradiation temperature (Figure \ref{fig:therm_pc}). Offsets steadily decrease from 55$^{\circ}$ to 12$^{\circ}$ and amplitudes steadily increase from 0.25 to 0.9 across irradiation temperature. With respect to that baseline, the inclusion of clouds dramatically increases the amplitude and decreases the offsets of cool, homogenously clouded planets. This is a natural consequence of starlight being absorbed higher in the atmosphere, where radiative timescales are shorter and there is less time for dynamics to redistribute heat before it is lost to space. For T$_\mathrm{irr} \geq 2000$\,K, clouds induce similar effects, but for different reasons. Cloudy nightsides are unable to cool as efficiently beneath the cloud deck, leading to the gas advected to the morning limb being hotter. Additionally, clouds present on the evening limb raise the thermal photosphere to cooler layers. Together, these work to decrease hotspot offsets and increase amplitudes \citep[as in][]{Parmentier2021grid, Roman2021}. This applies in our models with or without the inclusion of magnetic effects.

The inclusion of magnetism has a qualitatively similar effect on the phase curves, as zonal drag on the dayside mutes the equatorial jet, hampering heat redistribution to the nightside (increasing amplitude) and decreasing the eastward shift of the hotspot (decreasing offset). As temperatures rise, magnetism becomes the dominant effect on the offset, but the reduction of nightside flux by clouds continue to significantly increase the amplitude. The offsets of the bolometric thermal phase curves tend toward $\sim12^{\circ}$ for non-dragged models and $\sim4^{\circ}$ in dragged models.

For cooler hot Jupiters (T$_\mathrm{irr} \leq 2000$\,K), extended clouds (and to a lesser extent compact clouds) reflect significant starlight. Due to roughly homogeneous cloud coverage over the planet, the disk-integrated reflected light is maximized when the disk is fully illuminated (i.e. an offset near zero). As temperatures rise, clouds on the dayside begin to dissipate eastward of the substellar point, leading to a westward-shifted (negative) offset. For T$_\mathrm{irr} \geq 2750$\,K, there are few clouds left to dissipate on the eastern dayside, so further temperature increases dissipate mostly western clouds, shifting the reflected maximum back toward the substellar point. In magnetically dragged models, the temperature structure of the dayside is more symmetric, so the asymmetries in cloud coverage are less pronounced, and maxima are shifted toward the substellar point. We note, however, that this would be difficult to observe, as the total reflection from clouds is fairly low and thermal emission can be non-negligible in the same wavelength for planets this hot. Additionally, the temperatures on the morning and evening limbs are slightly variable in the cloudy, magnetic model cases for T$_\mathrm{irr} \geq 2500$\,K. This leads to corresponding changes in cloud coverage and a slightly variable reflected light offset. The fact that this variability only occurs in this model case is interesting, but more detailed analysis is left to future work.

\section{Discussion}
\label{sec:discussion}

Our results show that there is mutual feedback between clouds and magnetic effects, across much of the hot Jupiter population. However, both of these processes are complex and we have made simplifying assumptions to lower computational cost and make this study feasible. Here we discuss some of the caveats and possible limitations of our work, both as pertains to clouds and magnetic effects, as well as radiative transfer and the energetic effects of H$_2$ dissociation and recombination.

\subsection{Cloud assumptions}
In the interest of computational efficiency, our base cloud prescription only takes into account local thermodynamic conditions, including neither microphysical cloud formation nor the 3-D transport of clouds. The maximum vertical extent of clouds, a prescribed parameter in our treatment, should in reality be set by a balance of particle size-dependent gravitational settling and vertical mixing. The strength of vertical mixing, typically parameterized in one dimension with the eddy diffusion coefficient K$_\mathrm{zz}$, is particularly difficult to estimate. Different modeling approaches yield estimates that vary by orders of magnitude \citep{Moses2013, Parmentier2013TiOcoldtrapping,Agundez2014, Lines2018cloudGCM, Powell2018}. \citet{Komacek2019mixing} developed an analytic theory for the strength of vertical mixing, aligned with values from detailed tracer mixing in a GCM, predicting stronger mixing in hotter planets. Additionally, clouds that condense deeper in the atmosphere should be more compact, due to larger particles settling more efficiently \citep{Lines2018cloudGCM}. Sedimentation efficiency contributes additional uncertainty to cloud thickness \citep{Christie2021mixing_sedimentation}. {2-D microphysical models show that in addition to vertical mixing, longitudinal transport can qualitatively change cloud distributions, homogenizing clouds across the atmosphere and increasing cloud formation efficiency significantly \citep{Powell2024}. Considering the 3-D transport of cloud material can complicate cloud distributions further, as localized up- or down-welling can cause cloud-forming material to be depleted or enhanced as a function of latitude, longitude, and pressure. This can lead to patchier clouds than predicted from the purely thermodynamic conditions considered here \citep[e.g.][]{Tan2021browndwarfcloud,Komacek2022patchyclouds}.} The base cloudy cases in this work do not apply a limit to vertical extent (equivalent to assuming very strong vertical mixing globally), making them a upper limit on the presence of clouds. This assumption is likely most appropriate for the high-temperature end of our grid, where cloud bases are higher in the atmosphere and K$_{zz}$ is predicted to be larger. Thus our base set of cloudy models may be overestimating the radiative impact of clouds, particularly on the cooler end of our grid, but our compact models provide an example of how changes to the vertical extent of the clouds may impact our findings.

We chose to exclude some potential cloud-forming species (notably, strongly absorbing Fe) based on the finding of \citet{Gao2020} that they face significant nucleation barriers that may limit their presence in hot Jupiter atmospheres. Of the clouds included here, reflective silicates dominate the optical properties of the clouds. This choice strongly affects the Bond albedos of our cool, cloudy models \citep[compare ``extended nucleation-limited" and ``extended" cases in Figure 2 of][]{Roman2021}. 
We also assume that the optical properties of mixed-composition clouds are the optical-depth weighted average of each component species. Physically, however, mixed-composition clouds may form with some species being coated entirely by others, changing the optical properties of the grain \citep{Helling2006dirtyclouds}. Similarly, the fraction of available cloud mass is allowed to condense (set to 10\% in this work) should in principle be set by microphysical processes not included in our prescription. Each of these assumptions have an impact on the strength of cloud radiative feedback, and secondarily cloud distributions due to resulting changes in the temperature structure.

\subsection{Magnetic assumptions}
Our parameterization for magnetic effects assumes a dipolar magnetic field generated in the planet's interior and aligned with its rotation axis. {We additionally assume that magnetic field generated by the ions in the atmosphere is a small perturbation the background field. This assumption is justified when the magnetic Reynolds number (R$_\mathrm{m}$) is much less than unity}, where 
\begin{equation}
  \text{R}_\mathrm{m}=\frac{H U}{\eta}  
\end{equation}
where H is the pressure scale height, U is the zonal wind speed, and $\eta$ is the magnetic resistivity. {For models with T$_{irr} \geq 2000$ K, R$_{\rm m} > 1$ for some portion of the modeled atmosphere, at first only marginally on the upper dayside, but eventually for the entire dayside and parts of the high-altitude nightside. More detailed MHD simulations of hot Jupiter atmospheres have shown that when R$_{\rm m} > 1$, the magnetic field evolves in time, deviates from a dipolar structure, and the effect on atmospheric flows consequently becomes more complex than a zonal drag. This can lead to time-variable winds, including retrograde (east-to-west) zonal-mean flows \citep{Rogers2014, Hindle2021MHD}. The transition from the strict R$_{\rm m} < 1$ regime to a R$_{\rm m} \gtrsim 1$ regime coincides with the irradiation temperature at which magnetic drag begins to impact global structure, underscoring the need for more complex treatments for magnetic effects in simulations of hot Jupiter atmospheres. This is computationally daunting, however, particularly given the additional need for coupling to realistic radiative transfer and cloud modules.
While the magnetic drag treatment for the hottest planets in our grid is likely insufficiently complex for the problem, within the intermediate range, where the mutual interaction between clouds and magnetism is the most significant, this should be less of a concern.} 
Additionally, while we find a clear transition at T$_\mathrm{irr} = $ 2000 K at which magnetic effects begin to matter, this carries with it the implicit assumption of a 3 G magnetic field strength. A field strength above or below this value would modify the irradiation temperature at which magnetic effects become important. 

{For simplicity, our magnetic drag treatment applies only to zonal flows. While to first order hot Jupiter dynamics are dominated by zonal flows, meridional flows are not negligible, particularly at lower pressures and also once we apply the drag on zonal flows. While we do not expect this assumption to impact the increased longitudinal symmetry in cloud coverage in hot, magnetically dragged models, it could qualitatively affect the differences magnetic drag induces in nightside cloud distributions, which are driven by heat transport over the poles in our models.}

\subsection{Picket-fence radiative transfer in hot models}
While picket-fence radiative transfer has been benchmarked in GCMs against a more complex correlated-k treatment for the test case of HD\,209458b (T$_\mathrm{irr} \sim 2000$\,K) \citep{Lee2021}, and tested in the RM-GCM with and without clouds for HD\,189733b (T$_\mathrm{irr} \sim 1700$\,K) and HD\,209458b \citep{Malsky2024}, this is the first time it has been applied to significantly hotter atmospheres in a GCM. While we do not have 1:1 correlated-k GCMs to compare to, models of the hottest planets in our sample produce substellar temperature profiles with deep inversions (between 0.1 and 1 bar), but roughly isothermal or weakly inverted profiles higher in the atmosphere. An isothermal upper atmosphere is not consistent with the dayside temperature profiles retrieved from observations of the hottest hot Jupiters \citep[e.g.][]{Evans2016,Cont2022,Coulombe2023,vanSluijs2023}. Inversions are induced by strong starlight absorbers, and are most often attributed to gas-phase TiO and VO. Under the assumption of equilibrium chemistry, these molecules can be present on hot Jupiter daysides for T$_\mathrm{irr}\gtrsim$ 1900 K \citep{Hubeny2003,Fortney2008}. The actual transition between TiO/VO-poor and TiO/VO-rich atmospheres is uncertain, due to condensation potentially trapping these molecules deep in the atmosphere \citep{Spiegel2009TiOcoldtrapping,Parmentier2013TiOcoldtrapping}. Observations indicate a later transition than expected from equilibrium chemistry \citep{Line2016, Mansfield2021}. 

The coefficients used in our model were derived assuming equilibrium chemistry, however, meaning depletion of TiO/VO due to condensation is not included in our treatment. As such, we would expect to see strongly inverted atmospheres \citep[as are produced in equilibrium-chemistry correlated-k GCMs in this irradiation regime, e.g.][]{Parmentier2018,Kreidberg2018,Tan2024GCMgrid,Roth2024grid}. While this suggests that the picket-fence approximation may not be performing as well for the upper atmospheres of hotter planets, which would have implications for emission spectra computed from these models, we do not expect this to impact our conclusions regarding the interactions between magnetic effects and clouds. In this regime, the dayside upper atmosphere is virtually cloud-free and dayside zonal winds are already magnetically dragged so heavily as to be negligible, so increasing the temperatures further would have little effect on either process.

\subsection{Rotation rate \& host spectral type}
The vast diversity of irradiation temperatures across the hot Jupiter population is only possible due to the variations in host spectral type (and therefore luminosity) and orbital period (and therefore rotation period). In the grid of models presented here, we chose to vary irradiation temperature independently of rotation rate to more easily disentangle the effects of increasing irradiation from other effects. For all models, our radiative transfer routine implicitly assumes a Sun-like spectral energy distribution. Increasing T$_\mathrm{irr}$ in our model at a fixed orbital (rotation) period, therefore, shifts the host spectrum up without changing its shape, which introduces some physical inconsistency. The rotation rate used for all models in this work (1.4 days) is most appropriate for a hot Jupiter orbiting a Sun-like star with an irradiation temperature of $\sim$2500 K, in the regime where the interactions explored in section \ref{sec:cloudmag} are maximized, so to retain a Sun-like host, in principle the period would be shorter for models hotter than this and longer for cooler models. While it is beyond the scope of this paper to repeat this study with consistent assumptions about stellar host and rotation period, we can point to other work that has explored this connection. 

In the ultra-hot Jupiter regime, adjusting rotation rate consistently with irradiation temperature in (non-magnetic) GCMs can weaken or eliminate the trend of increasing day-night contrast with increasing irradiation \citep{Tan2024GCMgrid}. In cooler hot Jupiters, consistently adjusting period with irradiation temperature retrieves the same qualitative trend in heat redistribution efficiency vs irradiation temperature found in fixed-rotation grids. The trend between between hotspot offsets and irradiation temperature becomes non-monotonic, however, in contrast to the steady decrease found in fixed-rotation grids \citep{Roth2024grid}. This is because the components of circulation that drive heat redistribution and hotspot offsets are affected differently by changing rotation rates.

The effects of changing the spectral type of the host is not as well-explored in GCMs. One-dimensional modeling, however, has been used to investigate the impact of host spectral type on the radiative equilibrium structures of hot Jupiters \citep{Molliere2015} and ultra-hot Jupiters \citep{Lothringer2019}, showing that cooler host stars produce more isothermal vertical structures in their planets' atmospheres, as the gap between wavelengths of starlight absorption and planetary emission shrinks. 

\subsection{H$_2$ dissociation \& recombination}
The RM-GCM does not currently include the effects of H$_2$ dissociation and recombination, predicted \citep{Bell2018} and observed \citep{Mansfield2020H2} to be significant in the hottest hot Jupiter atmospheres. In these atmospheres, H$_2$ will thermally dissociate in hotter regions (taking energy) and recombine once advected to cooler regions (releasing energy), making heat redistribution more efficient. \citet{Tan2019UHJcirc} ran a grid of GCMs across equilibrium temperatures from 1600 K to 3600 K (2250 K to 5000 K in T$_\mathrm{irr}$) with and without the effects of H$_2$ dissociation and recombination. 
They find that while the hotspot is spread out over the dayside by H$_2$ dissociation and recombination for T$_\mathrm{irr} \geq$ 2800 K, day-night heat redistribution is only modified for T$_\mathrm{irr} \geq$ 3400 K. In our grid, this means that the dayside temperature structure for T$_\mathrm{irr} \geq$ 2750 K could be modified, but day-night temperature contrast should not be significantly altered for any of our models. A more spread-out hotspot in our models would likely lead to more uniformly cloud-free daysides, potentially diminishing the difference in morning limb cloud coverage between magnetic and non-magnetic model cases discussed in section \ref{sec:cloudmag}.

\section{Conclusion}
\label{sec:conclusions}
In this work, we present a set of hot Jupiter GCMs spanning most of the irradiation range of the observed population. We ran GCMs with and without both radiatively active cloud formation and magnetism at each irradiation level to investigate how these processes operate together across the population. Our key conclusions are:
\begin{itemize}
    \item Cloud coverage is globally uniform in our extended cloudy models for T$_\mathrm{irr} \leq$ 1750 K. In this range, magnetic effects are negligible, and assumptions about vertical cloud extent have a large impact. At these temperatures, clouds have a strong effect on the global energy balance, with Bond albedos between 0.15 and 0.7, depending on the prescribed maximum cloud thickness. 
    \item At T$_\mathrm{irr} \approx$ 2000 K, clouds begin to dissipate on the dayside, causing a sharp decrease in Bond albedo. This coincides with the temperature at which magnetic drag becomes dynamically significant in both cloudy and clear models. Dayside clouds continue to increase the Bond albedo to A$_\mathrm{B} \geq 0.1$ for T$_\mathrm{irr} \lesssim 2500$\,K, regardless of the inclusion of magnetic effects.
    \item For T$_\mathrm{irr} \gtrsim$ 2000 K, magnetic effects act in concert with cloud formation. Magnetic models produce cloud distributions that are more symmetric about the substellar point on the dayside. Cloud distributions are also more equatorially concentrated compared to non-dragged counterparts, both on the nightside and on the morning limb. 
    \item These processes operate to lower emitted phase curve offsets and increase amplitudes, both individually and in tandem. Clouds dominate the changes to emitted light for T$_\mathrm{irr} \leq 1750$\,K. For T$_\mathrm{irr} \geq 2000$\,K, both clouds and magnetic effects are important in shaping the emitted phase curve. For T$_\mathrm{irr} \geq 2500$\,K, reflected light from clouds is more longitudinally symmetric in magnetic models due to decreased asymmetries in temperature.
\end{itemize}

While we find interesting interactions between clouds and magnetic effects in this grid, there are many other planetary parameters and physical processes that are not explored in this work. Bulk composition (metallicity), rotation rate, surface gravity, disequilibrium chemistry, formation of photochemical hazes, and the energetic effects of H$_2$ dissociation, to name a few, are all expected to be important in setting atmospheric conditions.
To understand these atmospheres in their full complexity, we will need to understand how all of these operate both individually and in tandem.
GCMs are a powerful tool for this purpose, and will continue to be a vital resource as we get better and better constraints on atmospheric properties with JWST and beyond.

\section{Data Availability}
The final atmospheric structure outputs from the GCM models presented here are made available to the community at https://doi.org/10.5281/zenodo.13942071 \citep{GCMoutputs}.

\section{Acknowledgements}
We thank the anonymous referee for their careful review, which has improved the quality of this paper.

This work was supported by NASA XRP grant 80NSSC22K0313 and a grant from the Simons Foundation (918227, Rauscher).
\newpage
\bibliography{main}{}
\bibliographystyle{aasjournal}


\section{Appendix}
Here we present a table of the parameters that ensure numerical stability, which are different for each T$_\mathrm{irr}$, but are held constant for every T$_\mathrm{irr}$ across different model cases. Additionally, we present maps of normalized temperature, wind patterns, and an indication of cloud coverage to supplement the information in Figure \ref{fig:summary}.

\begin{deluxetable}{lcccc}[h]
{\tablehead{\multicolumn{5}{c}{Variable Grid Parameters}}}
\startdata
\multicolumn{1}{c}{T$_\mathrm{irr}$ (K)} & \multicolumn{1}{c}{DTSPD} & \multicolumn{1}{c}{t$_\mathrm{rad}$} & \multicolumn{1}{c}{t$_\mathrm{sponge}$} & \multicolumn{1}{c}{t$_\mathrm{diss}$} \\
\hline
\hline
1000          & 4800 & 4     & 0.05         & 0.025      \\
1250          & 4800 & 4     & 0.05         & 0.025      \\
1500          & 4800 & 4     & 0.01         & 0.025      \\
1750          & 4800 & 4     & 0.01         & 0.025      \\
2000          & 4800 & 4     & 0.01         & 0.005      \\
2250          & 6000 & 4     & 0.005        & 0.005      \\
2500          & 6000 & 4     & 0.005        & 0.005      \\
2750          & 8000 & 4     & 0.005        & 0.005      \\
3000          & 8000 & 2     & 0.005        & 0.005      \\
3250          & 8000 & 2     & 0.001        & 0.001     
\enddata
\label{tab:appendix_table}
\caption{The numerical parameters used across our grid of GCMs to ensure stability. In column order, these are: dynamical time steps per planet day (DTSPD), radiative time step (in units of the dynamical timestep), sponge layer Rayleigh drag timescale (t$_\mathrm{sponge}$), and the hyperdissipation timescale (t$_\mathrm{diss}$), both in units of planet days.}
\end{deluxetable}

\begin{figure*}[h]
    \centering
    \includegraphics[width=\textwidth]{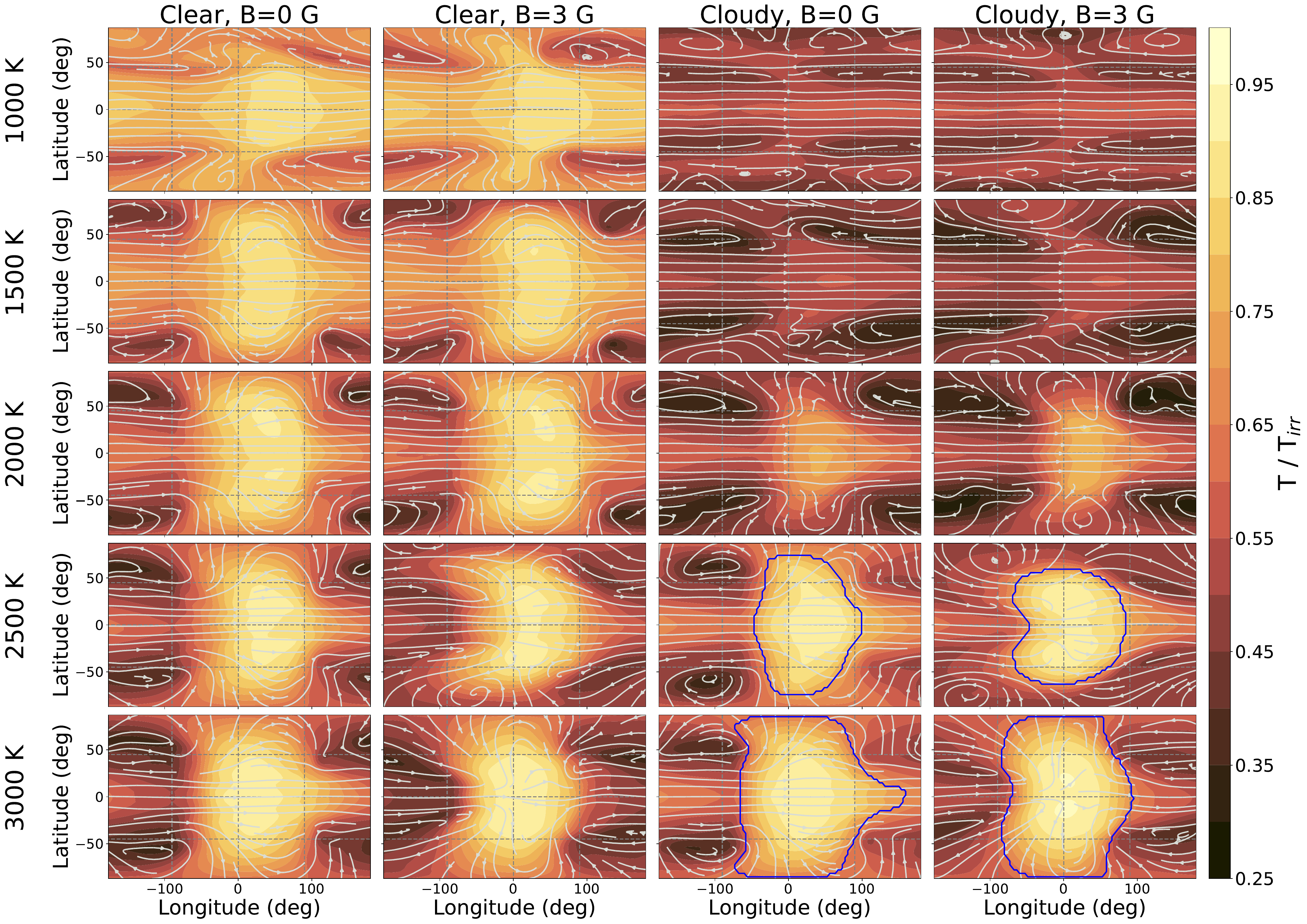}
    \caption{Normalized temperature (T/T$_\mathrm{irr}$) and wind maps at the 0.1 bar pressure level as in Figure \ref{fig:summary}, but for the irradiation temperatures not included in that figure. The blue contour separates regions where IR photospheres are cloud-free from those where clouds alter the depth of the IR photosphere. Note that the differences between the clear models with and without magnetism in the coolest model case (T$_\mathrm{irr} = 1000$\,K) are due to intrinsic time variability in the clear models, and are not associated with magnetic effects.}
    \label{fig:appendix_summary}
\end{figure*}

\end{document}